\documentclass[%
superscriptaddress,
reprint,
 amsmath,amssymb,
 aps,
pra,
nofootinbib
]{revtex4-1}

\usepackage{graphicx}
\usepackage[percent]{overpic}
\usepackage{dcolumn}
\usepackage{bm}
\usepackage{hyperref}
\usepackage{color} 
\definecolor{rltred}{rgb}{0.75,0,0}
\definecolor{rltgreen}{rgb}{0,0.6,0}
\definecolor{rltblue}{rgb}{0.3,0.3,1}

\usepackage{placeins}
\hypersetup{colorlinks,%
    hypertexnames=true,%
    linkcolor=rltred,%
    citecolor=rltgreen,%
    linktocpage=true}
    
\usepackage[normalem]{ulem} 
    
\begin{document} 

\title{Accurate high-harmonic spectra from time-dependent two-particle reduced density matrix theory}

\author{Fabian Lackner}
\email{fabian.lackner@tuwien.ac.at}
\affiliation{Institute for Theoretical Physics, Vienna University of Technology,
Wiedner Hauptstra\ss e 8-10/136, 1040 Vienna, Austria, EU} 

\author{Iva B\v rezinov\'a}
\email{iva.brezinova@tuwien.ac.at}
\affiliation{Institute for Theoretical Physics, Vienna University of Technology,
Wiedner Hauptstra\ss e 8-10/136, 1040 Vienna, Austria, EU} 

\author{Takeshi Sato}
\affiliation{Photon Science Center, School of Engineering, The University of Tokyo,7-3-1 Hongo, Bunkyo-ku, Tokyo 113-8656, Japan} 
\affiliation{Department of Nuclear Engineering and Management, School of Engineering, The University of Tokyo, 7-3-1 Hongo, Bunkyo-ku, Tokyo 113-8656, Japan} 

\author{Kenichi L. Ishikawa}
\affiliation{Photon Science Center, School of Engineering, The University of Tokyo,7-3-1 Hongo, Bunkyo-ku, Tokyo 113-8656, Japan} 
\affiliation{Department of Nuclear Engineering and Management, School of Engineering, The University of Tokyo, 7-3-1 Hongo, Bunkyo-ku, Tokyo 113-8656, Japan} 

\author{Joachim Burgd\"orfer}
\affiliation{Institute for Theoretical Physics, Vienna University of Technology,
Wiedner Hauptstra\ss e 8-10/136, 1040 Vienna, Austria, EU}  

\date{\today}

\begin{abstract}
The accurate description of the non-linear response of many-electron systems to strong-laser fields remains a major challenge. Methods that bypass the unfavorable exponential scaling with particle number are required to address larger systems. In this paper we present a fully three-dimensional implementation of the time-dependent two-particle reduced density matrix (TD-2RDM) method for many-electron atoms. We benchmark this approach by a comparison with multi-configurational time-dependent Hartree-Fock (MCTDHF) results for the harmonic spectra of beryllium and neon. We show that the TD-2RDM is very well-suited to describe the non-linear atomic response and to reveal the influence of electron-correlation effects. 
\end{abstract}

\maketitle

\section{Introduction} \label{sec:intro}
Solving the time-dependent Schr\"odinger equation for driven many-electron systems poses, to date, a major challenge. The long-range Coulomb interactions induce distant multi-particle correlations precluding the application of many approximations relying on short-ranged interactions from the outset. Numerically exact solutions have become available only for small systems, notably two-electron systems such as helium \cite{ParSmyTay98, PalResMcC08, FeiNagPaz08, HocBon11, PazFeiNag12} or H$_2$ \cite{AwaVanSae05, HarKonFuj02, VanHorMar06, SanPalCar07, PalBacMar07, GuaBarSch10, DehBanKam10}. As the numerical effort grows factorially with the number of electrons, an analogous treatment for large systems becomes prohibitive. Therefore, alternative strategies and approximations, most frequently the single-active electron approximation (SAE) \cite{KraSchKul92,Kul87} are invoked. For large systems time-dependent density functional theory (TDDFT) and the time-dependent Hartree-Fock (TDHF) approximation (e.g. \cite{ParYan89,Ull12}) have proven to provide an effective approach to electron dynamics on the mean-field level. The efficiency of TDDFT, however, comes with the price of introducing exchange-correlation functionals which are only known approximately and hard to improve systematically. Moreover, read-out functionals for two-particle observables are still largely unknown \cite{LapLee98, RohSimBur06, WilBau07}. Extension to the direct solution of the $N$-electron Schr\"odinger equation beyond the two-particle problem employs the multi-configurational time-dependent Hartree-Fock (MCTDHF) method (\cite{CaiZanKit05, AloStrCed07,HocHinBon14}). In principle, the MCTDHF method converges to the numerically exact solution if a sufficient number of orbitals is used. However, the factorial scaling with the number of particles limits its applicability. A recently proposed variant, the time-dependent complete active space self-consistent field (TD-CASSCF) method \cite{SatIsh13, SatIshBre16} which, in analogy to its ground-state counterpart, decomposes the state space into frozen, dynamically polarized, and dynamically active orbitals can considerably reduce the numerical effort yet eventually still leads to a factorial scaling with the number of active electrons $N^{\star}$ ($N^\star<N$). Further reduction of the numerical effort and, consequently, extension to larger systems appears possible, e.g., by applying the TD-ORMAS method \cite{SatIsh15}. 

The present approach intends to bridge the gap between full wavefunction-based $N$-electron descriptions such as MCTDHF and the time-dependent reduced one-particle density $\rho(\mathbf{r},t)$ based TDDFT. The underlying idea is to strike a compromise between accuracy of electron-electron correlations achieved by wavefunction-based methods and the ease to treat larger and, eventually, extended systems afforded by density-based approximations \cite{HaeCaiBou10}. Our working variable is the time-dependent two-particle reduced density matrix (2RDM), $D(\mathbf{r_1},\mathbf{r_2};\mathbf{r_1'},\mathbf{r_2,'};t)$. As a hybrid between the electron density and the full wavefunction the 2RDM contains the complete information on two-particle correlations but still scales polynomially with particle number \cite{ColYuk00, Maz07, Cio00}. Starting with the pioneering work in the 1950s \cite{Low55, Bop59}, the stationary 2RDM method for the ground state has matured to accuracies that often outperform those of coupled-cluster singles doubles with perturbative triples at similar or smaller numerical cost (see, e.g., \cite{NakNakEha01, NakEhaNak02, ZhaBraFuk04, Maz04,Maz06}). Extensions to the propagation of time-dependent systems have been pursued since the 90s \cite{SchReiToe90, GheKriRei93,TohSch14, BunNes08, AkbHasRub12, Bon98}. They have been plagued, however, by numerical instabilities which could be traced back to the approximations invoked for the three-particle reduced density matrix (3RDM). In a previous publication \cite{LacBreSat15} we have developed a TD-2RDM theory that is based on a contraction consistent reconstruction of the 3RDM, as required for the proper closure of the equations of motion for the TD-2RDM. It ensures that all constants of motion associated with symmetries of the Hamiltonian are conserved during time propagation. Contraction consistency of the reconstruction leads to a significant increase in the accuracy of the reconstruction, thereby achieving an accurate and stable propagation.

In the present paper we apply the TD-2RDM method to the non-linear response of many-electron atoms to strong few cycle laser pulses. The four electrons of beryllium and the ten electrons of neon are treated in their full dimensionality. We benchmark the resulting high-harmonic generation (HHG) against accurate spectra obtained by MCTDHF. HHG can be qualitatively well captured by the so called three-step model \cite{KraSchKul92, Cor93} in which an electron is first tunnel-ionized by the strong field, then accelerated in the laser field, and finally radiatively recombines with the parent ion emitting an energetic photon. For an accurate quantitative description it is crucial to explore and include correlation and many-electron effects neglected by such a one-electron model. In particular, the collective polarization response of the residual $N-1$-electron system to the external field and the ionized electron as well as the relaxation of the excitonic electron-hole pair are expected to be sensitive to correlation effects. We show that the TD-2RDM method is well suited to account for these subtle many-body effects and is superior to time-dependent mean-field descriptions such as TDHF or TDDFT. Going beyond our previous work \cite{LacBreSat15}, we include in the collision operator diagrammatic corrections of second order in line with the suggestions by Nakatsuji and Yasuda \cite{YasNak97} and Mazziotti \cite{Maz00_complete}. These second-order corrections turn out to be crucial for the accurate description of correlated HHG.

The paper is organized as follows. In Sec.~\ref{sec:theory} we briefly review key ingredients of the TD-2RDM theory. In Sec.~\ref{sec:methods} we summarize the methodological advances for reconstruction of the collision operator and purification of the 2RDM, essential ingredients for an accurate and stable propagation of the TD-2RDM. First applications of the TD-2RDM theory to full three-dimensional (3D) Be and Ne atoms are presented in Sec.~\ref{sec:results}. We show the time-dependent dipole moment and HHG spectra for the interaction with few-cycle laser pulses with varying intensities. We also compare to TDHF and TDDFT calculations. In Sec.~\ref{sec:summary} we give a short conclusion and outlook to future developments. Additional technical details are given in Appendices \ref{sec:appendix/CC}, \ref{sec:appendix/orbital_exp}, and \ref{sec:appendix/init}.  
%
\section{Brief review of the TD-2RDM method} \label{sec:theory}
The $p$-particle reduced density matrix ($p$RDM) $D(\mathbf{x}_1\dots\mathbf{x}_p;\mathbf{x}'_1\dots\mathbf{x}'_p;t)$ is defined by the trace over all but $p$ particles of the bilinear form $\Psi^\star\Psi$ of the $N$-electron wavefunction $\Psi$,
\begin{align}\label{eq:prdm}
&D(\mathbf{x}_1\dots\mathbf{x}_p;\mathbf{x}'_1\dots\mathbf{x}'_p;t) =  \nonumber\\ 
&\frac{N!}{(N-p)!} \int \Psi(\mathbf{x}_1\dots \mathbf{x}_p,\mathbf{x}_{p+1} \dots \mathbf{x}_N,t) \nonumber \\
&\times \Psi^*(\mathbf{x}'_1\dots \mathbf{x}'_p,\mathbf{x}_{p+1} \dots \mathbf{x}_N,t) \text{d}\mathbf{x}_{p+1}...\text{d}\mathbf{x}_N,
\end{align}
where $\mathbf{x}_i=(\mathbf{r}_i,\sigma_i)$ comprises the 3D space coordinate $\mathbf{r}_i$ and the spin coordinate $\sigma_i \in \{\uparrow,\downarrow\}$. With this definition the $p$RDM is normalized to $\frac{N!}{(N-p)!}$ which is a convenient choice for most calculations. Following \cite{Bon98}, we use for the $p$RDM the following short-hand notation
\begin{align}\label{eq:2rdm_short}
	D_{1\dots p} = D(\mathbf{x}_1\dots\mathbf{x}_p;\mathbf{x}'_1\dots\mathbf{x}'_p;t),
\end{align}
where also the time dependence in the notation of the RDMs is dropped for simplicity. 
The significance of the $p$RDM is that it allows the calculation of all joint $p$-particle observables. The RDMs thus contain in condensed form the full information on the $p$-particle correlations. This reduction of complexity compared to the $N$-electron wavefunction $\Psi$ is the key aspect making the propagation of the 2RDM so attractive and potentially useful for applications. In particular, the 2RDM carries the information on electron correlation at the two-particle level and thus, on total energy and two-particle excitation probabilities. The use of any exchange-correlation functional is avoided.\\
The electron density $\rho(\mathbf{r})$, and the pair density $\rho(\mathbf{r_1},\mathbf{r_2})$ are the diagonal elements of the 1RDM $D_1$,
\begin{align}\label{eq:rho1}
	\rho(\mathbf{r_1})=\sum_{\sigma}D(\mathbf{r_1}\sigma;\mathbf{r_1}\sigma)
\end{align}
and of the 2RDM $D_{12}$,
\begin{align}\label{eq:rho2}
	\rho(\mathbf{r_1},\mathbf{r_2})=\sum_{\sigma, \sigma'}D(\mathbf{r_1}\sigma\mathbf{r_2}\sigma';\mathbf{r_1}\sigma\mathbf{r_2}\sigma'),
\end{align}
measuring the probability to find simultaneously one of the particles at $\mathbf{r_1}$ and another one at $\mathbf{r_2}$. \\
The equations of motion of the RDMs belong to the Bogoliubov-Born-Green-Kirkwood-Yvon(BBGKY) hierarchy frequently invoked in classical and quantum statistical mechanics. The first two equations of the hierarchy read
\begin{align}
i\partial_tD_{1} &= \left[h_{1},D_{1}\right] + {\rm Tr}_2\left[W_{12},D_{12}\right],\label{eq:eom_d1}\\
i\partial_tD_{12} &= \left[H_{12},D_{12}\right] + {\rm Tr}_3\left[W_{13}+W_{23},D_{123}\right] \label{eq:eom_full_d12},
\end{align}
where $H_{12}$ is the reduced two-particle Hamiltonian,
\begin{align}\label{eq:2p_ham}
	H_{12} =h_1+h_2+W_{12},
\end{align}
consisting of the one-particle part $h_i$ containing the kinetic energy, the ionic and the external potential $V(\mathbf{r},t)$, and the electron-electron interaction $W_{12}$. The time-derivative of the $p$RDM depends on the next higher $(p+1)$RDM via the $p$-particle collision operator $C_{1...p}$. The lowest two members of the hierarchy are 
\begin{align}
    C_{1}\{D_{12}\} &={\rm Tr}_2\left[W_{12},D_{12}\right] \label{eq:C1},\\
	C_{12}\{D_{123}\} &={\rm Tr}_3\left[W_{13}+W_{23},D_{123}\right] \label{eq:C12}.
\end{align}
The coupling to higher-order members of the hierarchy makes the complete solution of the equations of motion as prohibitively complicated as the original TDSE for the $N$-particle problem. Therefore, any useful application invokes closure, i.e., approximating higher-order members of the $p$RDM hierarchy through functionals of lower-order members. In this work we close the equations by approximating the 3RDM in terms of the 2RDM
\begin{align}\label{eq:recon}
	D_{123}\approx D^{\rm R}_{123}\{D_{12}\}.
\end{align}
We will refer to this approximation as the reconstruction functional of the 3RDM. The equation of motion to be solved is thus 
\begin{align}
i\partial_tD_{12} &= \left[H_{12},D_{12}\right] + C_{12}\{D^{\rm R}_{123}\} \label{eq:eom_d12}.
\end{align}
Recently we have shown \cite{LacBreSat15} that contraction consistency
\begin{align}\label{eq:CC}
	D_{12}=\frac{1}{N-2}{\rm Tr}_3 D^{\rm R}_{123}\{D_{12}\},
\end{align}
of such a reconstruction $D^{\rm R}_{123}$, i.e., that the 2RDM used as the input to the reconstruction must be recovered after tracing out the coordinate of the third particle [Eq.~(\ref{eq:CC})], is of central importance for an accurate and consistent closure of the equation of motion for the 2RDM. It assures that constants of motion originating from symmetries of the Hamiltonian are conserved during propagation. Important examples are energy conservation for a time-independent Hamiltonian, and spin conservation for a spin-independent Hamiltonian. In fact, proper spin conservation requires further contraction relations given in Appendix~\ref{sec:appendix/CC}. Equation~(\ref{eq:CC}) ensures the contraction consistency of the first two members of the BBGKY-hierarchy, Eqs.~(\ref{eq:eom_d1}) and~(\ref{eq:eom_d12}), i.e., the second equation in the BBGKY-hierarchy, Eq.~(\ref{eq:eom_d12}), reduces to the first equation, Eq.~(\ref{eq:eom_d1}), when taking the trace over the second particle.

\section{Methodical developments} \label{sec:methods}

\subsection{Reconstruction} \label{sec:methods/reconstruction}
\begin{figure}
\centering
\includegraphics[width=1\linewidth]{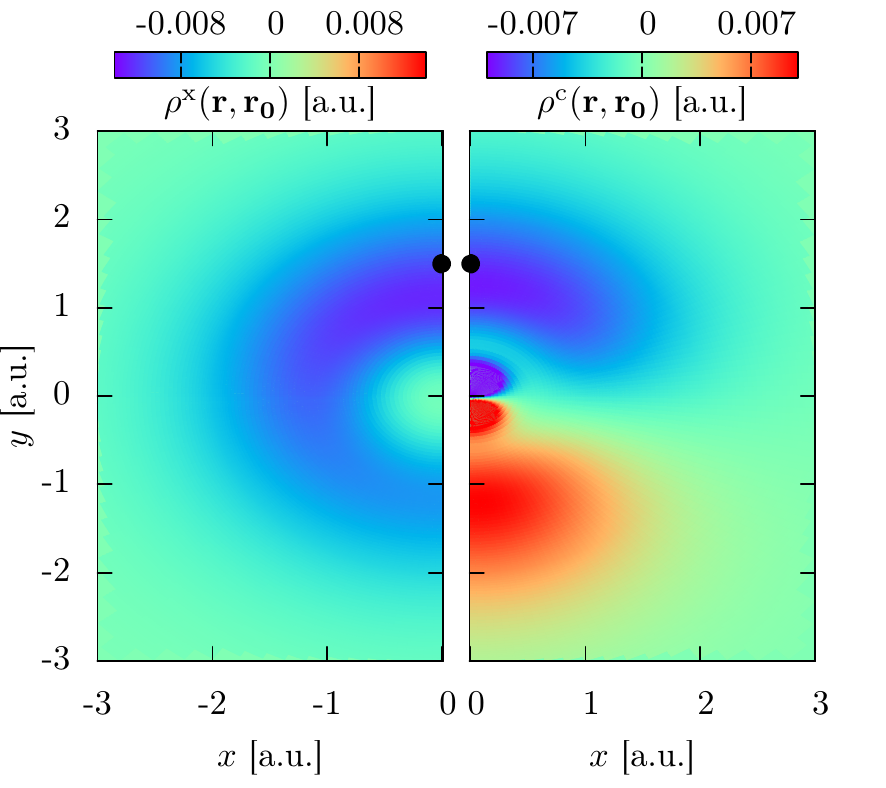}
\caption{(Color online) Exchange hole $\rho^{\rm x}(\mathbf{r},\mathbf{r_0})$ (left) [Eq.~(\ref{eq:rhoHF})] and Coulomb correlation hole $\rho^{\rm c}(\mathbf{r},\mathbf{r_0})$ (right) [Eq.~(\ref{eq:rhofull})] in the ground state of Be created by an electron located at the fixed position $\mathbf{r_0}=(0,1.5,0)$ (black dot). For a pure Slater determinant the exchange hole would be identical to the 2s orbital of the same spin. The asymmetry of the exchange hole originates from contributions beyond the single Slater determinant. The Coulomb correlation hole accounting for the electron repulsion is given by the two-particle cumulant [Eq.~(\ref{eq:rhoc})] and reduces the probability to find the second electron near the first one repelling the second electron to the other side of the atom.}
\label{fig:ground}
\end{figure}
For uncorrelated particles the 3RDM is given by the antisymmetrized tensor product of the 1RDM:
\begin{align}\label{eq:3rdmHF}
D^{\rm HF}_{123}=6\hat{A}D_1D_2D_3,
\end{align}
where $\hat{A}$ is the antisymmetrization operator normalized to be idempotent and the factor 6 originates from the number of distinct contributions. Equation~(\ref{eq:3rdmHF}) is therefore referred to as the Hartree-Fock approximation to the 3RDM, $D^{\rm HF}_{123}$. For correlated particles corrections to Eq.~(\ref{eq:3rdmHF}) are crucial. We introduce the corrections in terms of the cumulants $\Delta_{1...p}$, starting with the 2RDM. Accordingly, $D_{12}$ is analogously expanded into an uncorrelated tensor product $2\hat{A}D_1D_2$ describing independent identical particles and a proper correlation term $\Delta_{12}$ originating solely from the Coulomb interaction between the electrons
\begin{align}\label{eq:cumul2}
D_{12}=2\hat{A}D_1D_2 +\Delta_{12} = D^{\rm HF}_{12} + \Delta_{12}.
\end{align}
Explicitly, the Hartree-Fock approximation to $D_{12}$ is given by 
\begin{align}\label{eq:anti}
& D^{\rm HF}_{12} = 2\hat{A}D_1D_2 \nonumber \\
&=D(\mathbf{x}_1;\mathbf{x}'_1)D(\mathbf{x}_2;\mathbf{x}'_2)-D(\mathbf{x}_2;\mathbf{x}'_1)D(\mathbf{x}_1;\mathbf{x}'_2).
\end{align}
The term $\Delta_{12}$, referred to as the two-particle cumulant, is a sensitive measure for electron correlation and vanishes if and only if the particles are uncorrelated. It should be noted that the so called "Pauli correlation" resulting from antisymmetrization, i.e. exchange, are already contained in the Hartree-Fock term $D^{\rm HF}_{12}$ and is thus not part of $\Delta_{12}$. This decomposition applied to the diagonal elements of $D_{12}$ allows the identification of the proper correlation contributions to the pair density. For independent and distinguishable particles the pair density is a simple product of two individual one-particle probabilities $\rho(\mathbf{r_1})\rho(\mathbf{r_2})$. For identical fermions the Pauli principle prohibits the coalescence of two particles with the same spin giving rise to the exchange hole $\rho^{\rm x}$ (see Fig.~\ref{fig:ground}) and a deviation from a simple product
\begin{align}\label{eq:rhoHF}
\rho^{\rm HF}(\mathbf{r_1},\mathbf{r_2})&=\rho(\mathbf{r_1})\rho(\mathbf{r_2})+\rho^{\rm x}(\mathbf{r_1},\mathbf{r_2})\nonumber \\
&=\rho(\mathbf{r_1})\rho(\mathbf{r_2})-\sum_{\sigma}\vert D(\mathbf{r_1}\sigma;\mathbf{r_2}\sigma)\vert^2.
\end{align}
The Coulomb interactions lead to the appearance of additional Coulomb correlation $\rho^{\rm c}$ in the pair density, 
\begin{align}\label{eq:rhofull}
\rho(\mathbf{r_1},\mathbf{r_2})=\rho^{\rm HF}(\mathbf{r_1},\mathbf{r_2})+\rho^{\rm c}(\mathbf{r_1},\mathbf{r_2}).
\end{align}
This contribution to the pair density (see Fig.~\ref{fig:ground}) is directly given by the two-particle cumulant 
\begin{align}\label{eq:rhoc}
\rho^{\rm c}(\mathbf{r_1},\mathbf{r_2})=\sum_{\sigma, \sigma'}\Delta(\mathbf{r_1}\sigma\mathbf{r_2}\sigma';\mathbf{r_1}\sigma\mathbf{r_2}\sigma').
\end{align}
Using the decomposition, Eqs.~(\ref{eq:rhoHF}) and (\ref{eq:rhofull}), the exact total electron-electron interaction energy,
\begin{align}\label{eq:inter}
E^{ee}=&\int \frac{\rho(\mathbf{r_1},\mathbf{r_2})}{\vert \mathbf{r}_1-\mathbf{r}_2 \vert }\text{d}\mathbf{r}_1\text{d}\mathbf{r}_2,
\end{align}
follows as a sum of Hartree, exchange and correlation energy. With the help of the cumulant expansion [Eq.~(\ref{eq:cumul2})] the correlation contribution to other observables can be identified analogously. \\
The two-particle cumulant $\Delta_{12}$ can now be used to improve the reconstruction of $D_{123}$ beyond the Hartree-Fock level. Inclusion of corrections to first order yields 
\begin{align}\label{eq:3rdm_val}
D_{123}^{\rm V} = D^{\rm HF}_{123}+9\hat{A}\Delta_{12}D_3,
\end{align}
referred to as the Valdemoro (V) reconstruction which has been derived using a variety of different techniques \cite{ColPerVal93}. Going beyond first order by including second- or higher-order corrections has remained a challenge. Currently no technique appears to be available that allows to calculate the second-order correction for all elements of the 3RDM in terms of the lower order RDMs. However, for specific elements second-order contributions can be calculated by following two different approaches developed by Nakatsuji and Yasuda \cite{YasNak97} and Mazziotti \cite{Maz00_complete}. \\
%
\begin{figure}
	\centering
	\includegraphics[width=1\linewidth]{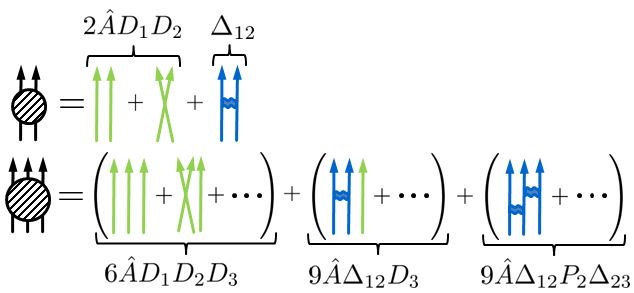}
	\caption{(Color online) Diagrammatic representation of the Nakatsuji-Yasuda reconstruction for the 3RDM, Eq.~(\ref{eq:3rdm_ny}), in analogy to the diagrammatic expansion of many-body Green's functions \cite{Mat76}. Green arrows correspond to one-particle propagators and blue parts describe connected diagrams accounting for particle-particle interactions. The brackets represent all topologically distinct permutations of a given diagram ensuring antisymmetry of the 3RDM.}
	\label{fig:diagrams}
\end{figure}\noindent
The Nakatsuji-Yasuda (NY) approach is based on the diagrammatic expansion of the 3RDM in analogy to many-body Green's functions (Fig.~\ref{fig:diagrams}). Their result for the second-order correction can be written as
\begin{align}\label{eq:3rdm_ny}
D_{123}^{\rm NY} &= D^{\rm HF}_{123}+9\hat{A}\Delta_{12}D_3+9 \hat{A} \Delta_{12}P_2\Delta_{23}\nonumber \\
&= D^{\rm V}_{123}+9 \hat{A} \Delta_{12}P_2\Delta_{23},
\end{align}
where the operator $P_i$ acting on the $i^{\rm th}$ particle is given by
\begin{align}\label{eq:T}
P_i=2D^{\rm HF}_i-I_i,
\end{align}
with the one-particle identity matrix $I_i$ and the reference Hartree-Fock 1RDM
\begin{align}\label{eq:T}
D^{\rm HF}(\mathbf{x}_1;\mathbf{x}'_1)=\sum_i^N f_i(\mathbf{x}_1)f_i^*(\mathbf{x}'_1),
\end{align}
constructed from the eigenfunctions of the 1RDM $f_i$, i.e. the natural orbitals. This present formulation differs from the original work by Nakatsuji and Yasuda who chose $f_i$ as the Hartree-Fock orbitals rather than the natural orbitals \cite{YasNak97}. The leading error in the NY-reconstruction is the neglect of second-order corrections for triple excitation amplitudes  $\Delta_{ooo}^{uuu}$, where $o$ denotes occupied and $u$ denotes unoccupied orbitals in the Hartree-Fock reference. \\
An alternative approach to second-order corrections introduced by Mazziotti \cite{Maz00_complete} yields an implicit equation for the second-order correction 
\begin{align}\label{eq:3rdm_maz}
\Delta^{\rm M}_{123}=\hat{A}D_1\Delta^{\rm M}_{123}+\hat{A}\Delta^{\rm M}_{123}D_1+3\hat{A}\Delta_{12}\Delta_{23},
\end{align}
entering the reconstruction of the 3RDM 
\begin{align}\label{eq:3rdm_maz_full}
D^{\rm M}_{123} = D^{\rm V}_{123} + \Delta^{\rm M}_{123}.
\end{align}
By transforming into the eigenbasis of the 1RDM Eq.~(\ref{eq:3rdm_maz}) can be explicitly solved except for elements $\Delta^{uuu}_{ooo}$ which stay undetermined \cite{Maz00_complete}.\\
All reconstruction functionals $D^{\rm R}_{123}$ as introduced above are not contraction consistent, i.e. do not fullfill Eq.~(\ref{eq:CC}). Previously, we developed a generic method to enforce contraction consistency for an arbitrary reconstruction functional. This allows us to extend the reconstruction functionals $D^{\rm V}_{123}$, $D^{\rm NY}_{123}$ and $D^{\rm M}_{123}$ to contraction consistent form. Our method is based on the unitary decomposition of the 3RDM \cite{LacBreSat15}
\begin{align}\label{eq:unita}
D_{123}=D_{123;\perp}\{D_{12}\}+D_{123;\rm K},
\end{align}
where the kernel $D_{123;\rm K}$ is defined as the component of $D_{123}$ whose contractions vanish, $\text{Tr}_3 D_{123;\rm K}=0$. Because the component perpendicular to the kernel $D_{123;\perp}\{D_{12}\}$ is a known functional of the 2RDM, only the kernel component needs to be approximated. To make a given reconstruction functional contraction consistent we use Eq.~(\ref{eq:unita}) and replace the unknown exact kernel component $D_{123;\rm K}$ by the kernel component of the chosen approximate reconstruction $D^{\rm R}_{123;\rm K}$. In the following, we will apply this contraction consistency constraint to $D^{\rm V}_{123}$, $D^{\rm NY}_{123}$ and $D^{\rm M}_{123}$.

\subsection{$N$-representability and purification} \label{sec:methods/purif}

The realization of the goal to replace the propagation of the $N$-particle wavefunction by that of the 2RDM faces, in addition to the reconstruction (or closure) problem, a second and closely intertwined conceptual hurdle, known as the $N$-representability problem. While $\Psi(t)$ itself is not needed at any time during the propagation of the equations of motion [Eq.~(\ref{eq:eom_d12})], the solution of $D_{12}(t)$ must, at all times, satisfy Eq.~(\ref{eq:prdm}), i.e. it must be representable as a partial trace over the bilinear form $\Psi^*(t) \Psi(t)$ of an (unknown) $N$-particle wavefunction. If the exact form of $D_{123}$ were to be used in Eq.~(\ref{eq:eom_d12}), this would be trivially the case. However, as soon as approximations to $D_{123}$ are employed, the time evolved $D_{12}(t)$ may leave the subspace of $N$-representable 2RDMs. While an explicit complete set of conditions for $N$-representability is still unknown several necessary conditions are well-established. Among those the D-condition and the Q-condition are the most important. They guarantee that the 2RDM as well as the two-hole reduced density matrix (2-HRDM)
\begin{align}\label{eq:HRDM}
Q_{12} &= 2 \hat{A}I_1 I_2-4 \hat{A}I_1 D_2 + D_{12},
\end{align}
remain both positive semidefinite (i.e.~have non-negative eigenvalues) denoted by 
\begin{align}
D_{12}& \geq 0, \label{eq:posD}\\
Q_{12}& \geq 0. \label{eq:posQ}
\end{align}
The positive semi-definiteness of these matrices represent mutually independent conditions although the matrices are inter-convertible by Eq.~(\ref{eq:HRDM}). The 2-HRDM describes the pair distribution of holes rather than of particles and its positivity in combination with the positivity of the 2RDM ensures that the occupation number of a particle pair or a hole pair in any two-body state is always non-negative.
Violation of the positive semi-definiteness obviously causes instabilities in the propagation which, in turn, are consequences of the approximation error in $D^{\rm R}_{123}$. To avoid such instabilities we enforce the D- and Q-condition during propagation by a purification process. Additional $N$-representability conditions, e.g. the G-condition were found to remain well satisfied if the D- and Q-condition are simultaneously enforced. \\
We briefly describe the present new purification scheme which improves the numerical efficiency compared to its predecessor \cite{Maz02Pur}. To isolate the defective part of the 2RDM we decompose the Hermitian 2RDM $D_{12}=D^{<}_{12}+D^{>}_{12}$ into components with negative eigenvalues $D^{<}_{12}$ and positive eigenvalues $D^{>}_{12}$,
\begin{align}
D^{<}_{12}=\sum_{g_i < 0} g_i \vert g_i \rangle \langle g_i\vert, \\
D^{>}_{12}=\sum_{g_i > 0} g_i \vert g_i \rangle \langle g_i\vert,
\end{align}
where $g_i$ are the eigenvalues and $\vert g_i \rangle$ are the eigenfunctions (called geminals) of the 2RDM. Simply neglecting $D^{<}_{12}$ in the decomposition of the 2RDM is not a viable option as it would lead to uncontrolled errors in the normalization as well as the associated 1RDM. Instead we employ the unitary decomposition of the 2RDM in analogy to Eq.~(\ref{eq:unita}) 
\begin{align}
D^{<}_{12}=D^{<}_{12;\perp}\{D_{12}\}+D^{<}_{12;\text{K}}.
\end{align}
As above, the kernel $D^{<}_{12;\text{K}}$ is by definition fully contraction free. Therefore, subtracting the kernel component from the 2RDM,
\begin{align}
D'_{12} = D_{12}-D^{<}_{12;\text{K}},
\end{align}
leaves the norm and the 1RDM invariant.
After a single purification step $D'_{12}$ has a significantly reduced negative eigenvalues. The same unitary decomposition can be applied to simultaneously enforce the approximate positivity of the 2-HRDM
\begin{align}\label{eq:cumul}
D'_{12} = D_{12}-D^{<}_{12;\text{K}}-Q^{<}_{12;\text{K}}.
\end{align}
Since the negative eigenvalues of the 2-HRDM have dominant contributions in the high occupation numbers of the 2RDM the two matrices $D^{<}_{12;\text{K}},Q^{<}_{12;\text{K}}$ act on different subspaces and, therefore, do not destroy the purifying effect of the other but lead to a simultaneous enforcement of the D- and Q-condition. The effectiveness of the purification is further improved by applying it iteratively.

\subsection{Probing for the reconstruction error} \label{sec:methods/recon_error}
\begin{figure}
	\centering
	\includegraphics[width=1\linewidth]{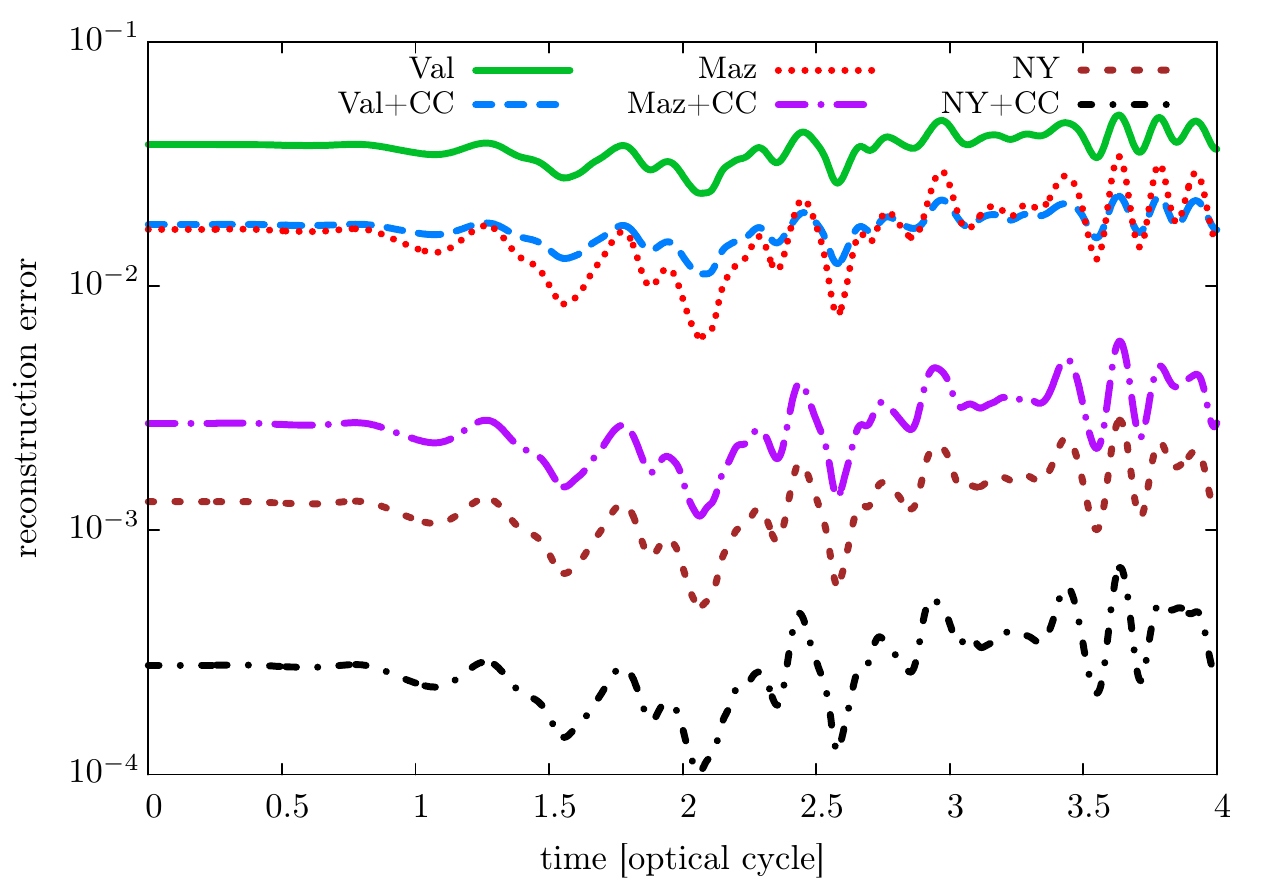}
	\caption{(Color online) Comparison of the distance [Hilbert-Schmidt norm Eq.~(\ref{eq:error})] between the exact and reconstructed 3RDM for different reconstruction functionals proposed by Valdemoro (V) [Eq.~(\ref{eq:3rdm_val}]), Nakatsuji-Yasuda (NY) [Eq.~(\ref{eq:3rdm_ny})] and Mazziotti (M) [Eq.~(\ref{eq:3rdm_maz})] with and without including contraction consistency denoted by CC [Eq.~(\ref{eq:CC})]. The simulation is performed for Be subject to a two-cycle laser pulse with $I=0.5\times 10^{14}\rm{W/cm^2}$ and $\lambda = 800\rm nm$ (for details see Sec.~\ref{sec:results}). The input 2RDM for the reconstruction is taken from an MCTDHF calculation to avoid the accumulation of propagation errors.}
	\label{fig:error}
\end{figure}
We test now different reconstruction functionals introduced in Sec.~\ref{sec:methods/reconstruction} since choosing the most accurate reconstruction is key to obtain highly accurate HHG spectra. Note that in this subsection we do not solve the closed equation of motion of the 2RDM [Eq.~(\ref{eq:eom_d12})]. Instead we use the exact 2RDM and 3RDM from an MCTDHF calculation to compare $D^{\rm R}_{123}$ [Eq.~(\ref{eq:recon})] reconstructed from the exact 2RDM with the exact 3RDM at each time step. This allows a direct measurement of the pure reconstruction error without admixture of any propagation error. We measure the distance between the reconstructed 3RDM $D^{\rm R}_{123}$ and the exact 3RDM $D_{123}$ by the square of the Hilbert-Schmidt norm for matrices,
\begin{align}\label{eq:error}
\| D_{123}-D^{\rm R}_{123} \|^2=\text{Tr}_{123}(D_{123}-D^{\rm R}_{123})^2.
\end{align}
We find (Fig.~\ref{fig:error}) that, as expected, the largest error occurs for the Valdemoro reconstruction which takes into account correlations only up to first order [Eq.~(\ref{eq:3rdm_val})]. Further improvement can be obtained by including second-order contributions. While the Mazziotti reconstruction  $D^{\rm M}_{123}$ considerably improves upon the Valdemoro reconstruction, the Nakatsuji-Yasuda reconstruction  $D^{\rm NY}_{123}$ [Eq.~(\ref{eq:3rdm_ny})] provides the most accurate results for the present atomic systems under investigation (Fig.~\ref{fig:error}). Irrespective of the perturbative order included we consistently find for all three reconstructions significant improvement in the accuracy of the reconstruction when enforcing contraction consistency [Eq.~(\ref{eq:CC})]. Contraction consistency is thus not only essential for preserving fundamental symmetries during propagation but also leads to a significantly better reconstruction. As an aside we note that this observation suggests that contraction consistency should improve also the solution of the anti-Hermitian contracted stationary Schr\"odinger equation used to calculate the ground state of molecules (see e.g.~\cite{Maz06_antiher})
for which the basic approximation is also the reconstruction of the 3RDM. \\
The remaining error in the reconstruction functional comes from additional second-order contributions not included in the Nakatsuji-Yasuda or Mazziotti approximation. Future work will address the systematic inclusion of all second-order and, possibly, higher-order corrections. In the following simulations we will apply the contraction-consistent Nakatsuji-Yasuda approximation.

\section{High-harmonic generation} \label{sec:results}
\begin{figure*}
	\centering
	\includegraphics[width=1\linewidth]{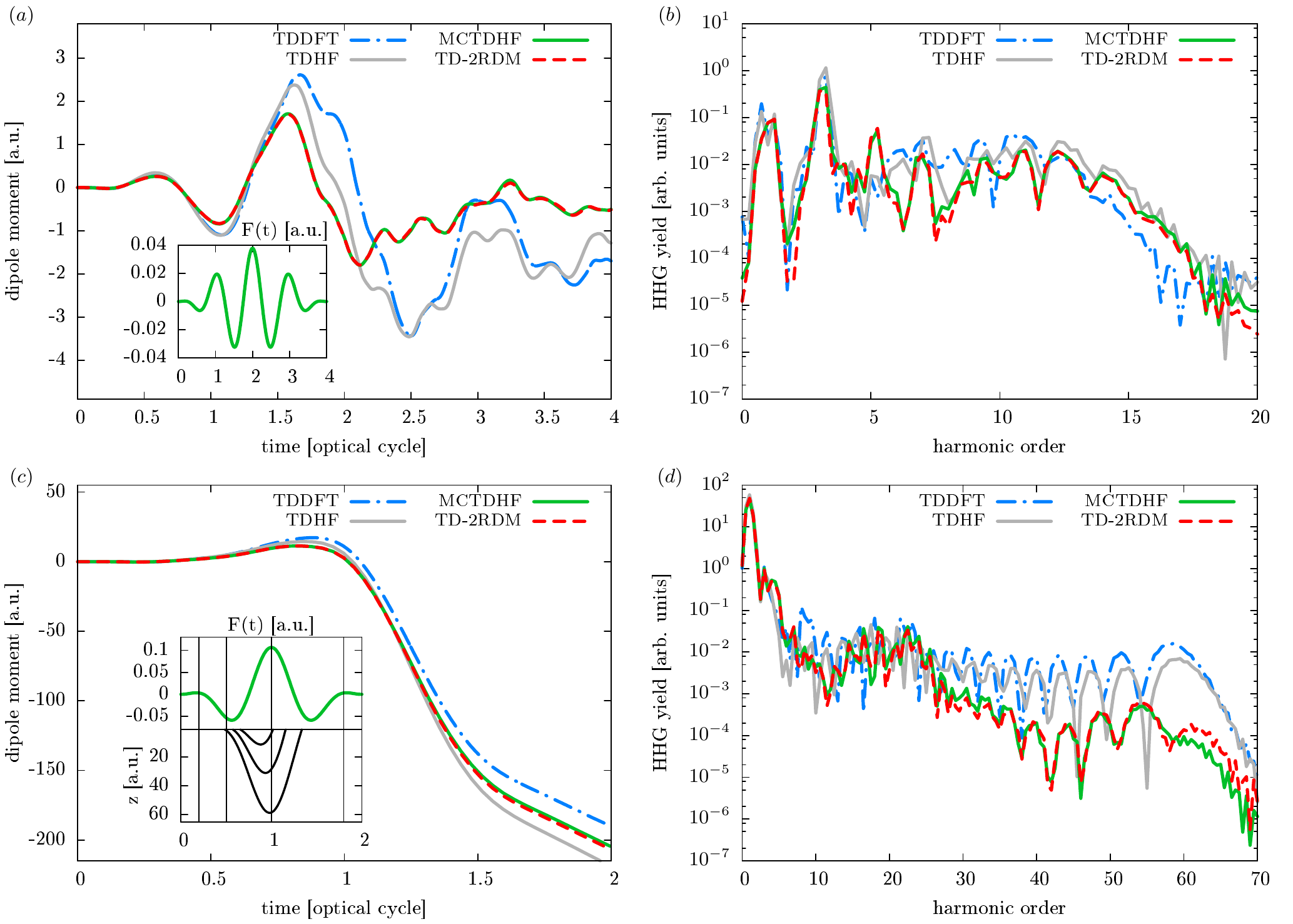}
	\caption{(Color online) Electronic response of Be subject to a four-cycle laser pulse with $I=0.5\times 10^{14}\rm{W/cm^2}$ [(a) and (b)] and a two-cycle laser pulse with $I=4.0\times 10^{14}\rm{W/cm^2}$ [(c) and (d)] monitored by the induced dipole moment [(a), (c)] and the high-harmonic spectrum [(b), (d)] obtained from the dipole acceleration [Eq.~(\ref{eq:lamor})]. The wavelength of the pulse is $\lambda = 800\rm nm$. The TD-2RDM result calculated with the contraction consistent Nakatsuji-Yasuda reconstruction is compared to MCTDHF, TDHF, and TDDFT within the local density approximation. The inset shows the electric field [Eq.~(\ref{eq:laser_pulse})] of the applied laser pulse with (a) four and (c) two cycles, respectively. The inset of (c) shows three representative electron trajectories, two with recollision energies contributing to the $30^{\rm th}$ harmonics and one (in the center) contributing to the high-energy cut-off within the three-step model. The vertical lines mark the four instances in time at which snapshots are taken in Fig.~\ref{fig:ph_correlation}. The TD-2RDM result for the stronger two-cycle laser pulse [(c), (d)] used a total number of 40 iterations per purification step whereas no purification was employed for the weaker pulse [(a), (b)].}
	\label{fig:multiplot_be}
\end{figure*}
In this section we present the first results of the TD-2RDM method as applied to full 3D multi-electron atoms. To benchmark the TD-2RDM theory we use a state of the art MCTDHF implementation published recently \cite{SatIshBre16} as well as TDDFT within the local density approximation with exchange functional \cite{Dir30} and correlation functional \cite{VosWilNus80} computed via the LIBXC library \cite{MarOliBur12}. We simulate the non-linear many-electron response of Be and Ne subject to ultra-short few-cycle laser pulses of the form 
\begin{align}\label{eq:laser_pulse}
F(t)=F_0 \cos(\omega t) \sin^2\left( \frac{\omega}{2N_c} t\right)  \quad 0 \le t \le N_c \frac{2\pi}{\omega},
\end{align}
where $F_0$ is the amplitude of the electric field, $\omega$ is the mean angular frequency, and $N_c$ is the number of cycles. We use from now on the scaled time $\tau=t  \frac{2\pi}{\omega}$ with $0 \le \tau \le N_c$, counting the fractional number of cycles that have passed.\\
The gauge invariance of the exact 2RDM equation of motion [Eq.~(\ref{eq:eom_full_d12})] is guaranteed by that of the underlying time-dependent Schr\"odinger equation. The approximate equation of motion [Eq.~(\ref{eq:eom_d12})] retains it if the employed reconstruction is invariant under unitary transformations of orbitals, which is the case for all the reconstructions discussed here. However, a specific numerical implementation is not necessarily gauge invariant. While the present implementation (see Appendix~\ref{sec:appendix/orbital_exp}) is gauge invariant, those with a finite number of time-independent spin orbitals are in general not. Due to the favorable numerical behavior \cite{SatIshBre16} we employ the velocity gauge in all our simulations. \\
We particularly focus on the time-dependent dipole moment and the HHG spectrum which depends sensitively on the electron dynamics and is, therefore, well suited to test the capabilities of the TD-2RDM method. In our simulations we consider a near-IR pulse with wavelength $\lambda = 800 \rm nm$ ($\omega = 0.057$ a.u.) which is the central wavelength of the Ti:sapphire laser.
%
We investigate the response of the Be atom for two laser pulses differing in intensity and in the number of cycles. One pulse has an intensity of $I=0.5\times 10^{14}\rm{W/cm^2}$ ($F_0=0.038$) and a length of four cycles $N_c=4$ showing a moderate amount of ionization. The second pulse with $I=4.0\times 10^{14}\rm{W/cm^2}$ ($F_0=0.107$) and two cycles leads to significant ionization [see the inset of Fig.~\ref{fig:multiplot_be} (a) and (c) for the pulse shape]. Because of the much larger ionization potential ($I_{\rm p}=21.6$ eV) we employ in our simulation for neon a stronger pulse with $I=10^{15}\rm{W/cm^2}$ ($F_0=0.169$). \\
The HHG spectrum $I^{\rm HHG}(\omega)$ is determined using the classical Lamor formula 
\begin{align}\label{eq:lamor}
I^{\rm HHG}(\omega)=\frac{2}{3c} \bigg\vert \int \ddot{d}(t) e^{i \omega t} \rm{d}t \bigg\vert^2,
\end{align}
where the expectation value of the dipole acceleration operator is calculated via 
\begin{align}\label{eq:dipole_acc}
\ddot{d}(t)&=-\langle \Psi(t) | \frac{\partial V(\mathbf{r})}{\partial \mathbf{r}} | \Psi(t) \rangle.\nonumber \\
&=-\int \frac{\partial V(\mathbf{r})}{\partial \mathbf{r}} \rho( \mathbf{r},t) \text{d}\mathbf{r}.
\end{align}
As the HHG spectrum is a functional of the reduced one-particle density $\rho( \mathbf{r},t) $, TDDFT can be directly applied without invoking any approximate read-out functional. \\
The numerical implementation of the TD-2RDM method requires the simultaneous solution of the orbital equations of motion Eq.~(\ref{eq:dphi_MCTDH}) in Appendix~\ref{sec:appendix/orbital_exp} together with the equation of motion for the 2RDM matrix elements Eq.~(\ref{eq:eom_approxspin}). We solve the angular part of the orbital equations of motion by expanding the orbitals in spherical harmonics with a maximum angular momentum of $L_{\rm max}=47$. The radial part of the orbital equations of motion are numerically solved by employing a finite-element discrete variable representation in radial direction with 50 finite elements and 20 basis points each. We use a total radial box size of 200 atomic units and absorbing boundary conditions implemented by a $\cos^{\frac{1}{4}}$ mask function starting at a radius of 160 atomic units with a transition length of 40 atomic units. To obtain converged results for the time-dependent dipole moment in the presence of electrons absorbed by the absorbing boundary we calculate the dipole moment by a double integral over time of the dipole acceleration. \\
The time propagation of the orbitals is performed using a second-order split-step method treating the one-body part $\hat{h}$ and the part containing the electron-electron interaction in Eq.~(\ref{eq:dphi_MCTDH}) separately. Further details on the propagation of the orbitals can be found in Ref.~\cite{SatIshBre16}. We apply this propagation in real and imaginary time, the latter for the determination of the MCTDHF ground state which also serves as the initial state for the TD-2RDM propagation. It should be noted that the "exact" MCTDHF ground state is, in general, not a stationary solution of the approximate equation of motion of $D_{12}$ [Eq.~(\ref{eq:eom_d12})] when a reconstruction approximation has been applied in the collision operator, $C_{12}\{D^{\rm R}_{123}\}$ [Eq.~(\ref{eq:recon})]. However, provided that the reconstruction approximation is sufficiently accurate, the stationary solution of Eq.~(\ref{eq:eom_d12}) is close to that of the MCTDHF solution and can be further improved upon by projecting out the residual excitations by time averaging over a field-free propagation (see Appendix~\ref{sec:appendix/init}). \\
We emphasize that the propagation of the TD-2RDM method is performed fully self-consistent using Eq.~(\ref{eq:eom_d12}) without invoking the many-body wavefunction at any time.

\subsection{Beryllium} \label{sec:results/beryllium}
As the first application we study the non-linear dipole response and HHG of Be for a laser pulse with $I=0.5\times 10^{14}\rm{W/cm^2}$. Due to the spherical symmetry of the ground state the initial dipole moment is zero. In the presence of the laser pulse the atom gets polarized and the dipole moment starts to oscillate [Fig.~\ref{fig:multiplot_be}(a)]. Initially the dipole moment nearly adiabatically follows the electric field of the external laser pulse but soon after the first cycle non-linear effects manifest themselves through the appearance of multiple frequencies. The near perfect agreement between the TD-2RDM method and MCTDHF shows that the TD-2RDM method is capable to accurately describe excitation processes in strong laser fields. In clear contrast, the mean-field TDDFT and TDHF methods show marked deviations during the entire pulse duration. In particular the overshoot of the first oscillations shows that the binding energy is underestimated within TDHF and TDDFT compared to MCTDHF. The TD-2RDM method on the other hand is capable of reproducing the initial oscillations, i.e. linear polarizability and binding energy, with high accuracy.
The high-harmonic spectrum which probes the frequency dependent electronic response yields excellent agreement between TD-2RDM and MCTDHF [Fig.~\ref{fig:multiplot_be}(b)]. TDHF and TDDFT reproduce the overall structure of the spectrum well but show deviations mostly in the plateau region [Fig.~\ref{fig:multiplot_be}(b)]. \\
%
\begin{figure}
	\centering
	\includegraphics[width=1\linewidth]{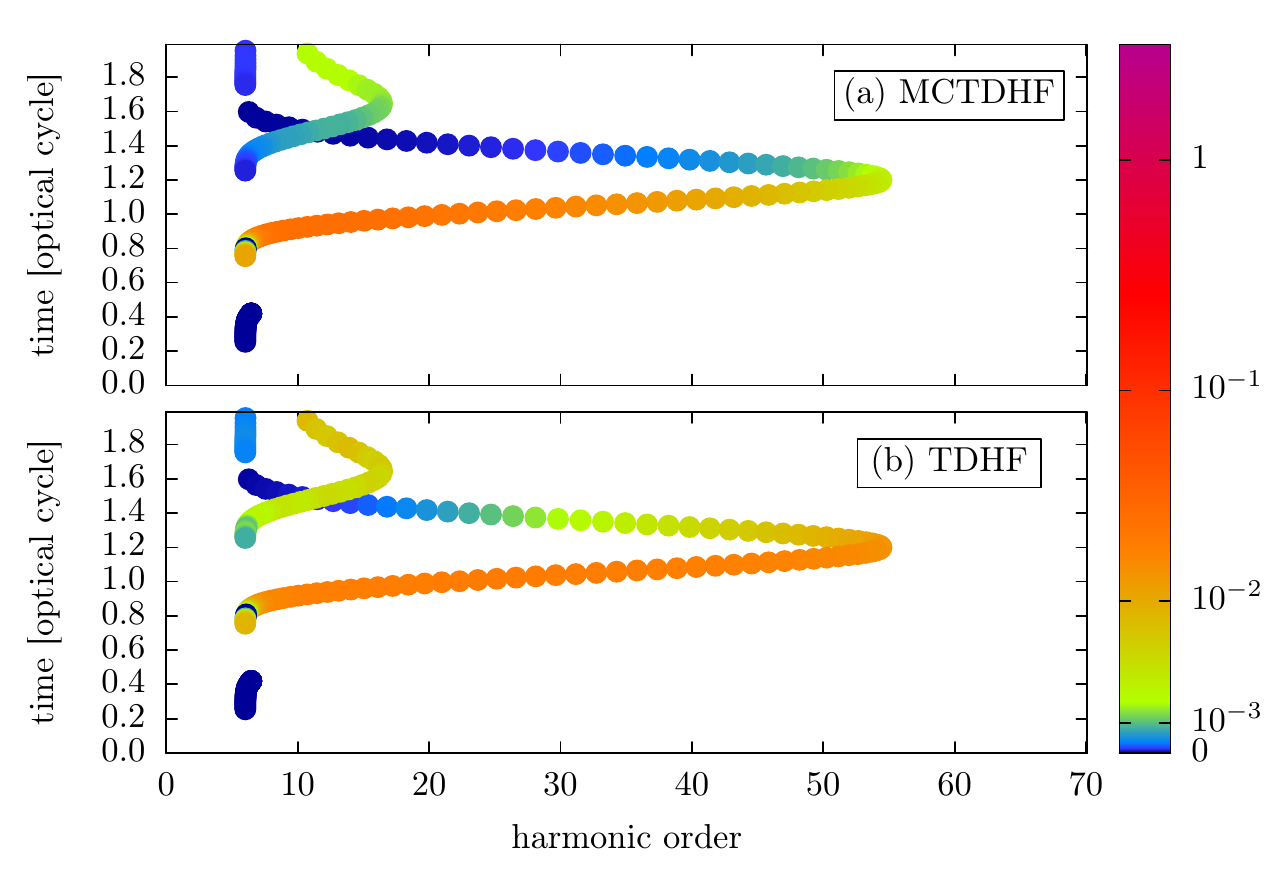}
	\caption{(Color online) Time-frequency spectrum of the high-harmonic radiation of Be subject to a two-cycle laser pulse with $I=4.0\times 10^{14}\rm{W/cm^2}$ and $\lambda = 800\rm nm$ predicted by the three-step model [Eq.~(\ref{eq:TSM_intensity})] using the ionization probabilities from (a) the TDHF or (b) the MCTDHF calculation as input. The absolute scale of the intensity has been adjusted for direct comparison to the result of the quantum mechanical calculation (Fig.~\ref{fig:STFT_be}).}
	\label{fig:TSM_be}
\end{figure}\noindent
%
The stronger pulse ($I=4.0\times 10^{14}\rm{W/cm^2}$) features already significant ionization reflected in the pronounced slope caused by the contribution of ionizing electrons [Fig.~\ref{fig:multiplot_be}(c)]. Unlike for lower intensities where the high-harmonic spectrum of TDHF and MCTDHF differ only in the detailed structure [Fig.~\ref{fig:multiplot_be}(b)] for this strong intensity we find a strong overestimate of the high-harmonic intensity predicted by TDHF [Fig.~\ref{fig:multiplot_be}(d)] previously reported in \cite{SatIshBre16}. This overestimate originates from the inability of mean-field theories such as TDHF or TDDFT to describe pure single electron ionization as TDHF and TDDFT inevitably introduce artificially enhanced double ionization whenever single ionization occurs \cite{HocBon11}. The, at first glance, somewhat counterintuitive consequence of the coupling between single and double ionization is that the overall probability for ionization occuring at all is reduced. This follows from the "contamination" of pure single ionization by ionization of the second more deeply bound electron with binding energy of the second ionization potential. In turn, the lower probability for ionization leads to a larger bound state component in the wavefunction of the residual ion enhancing the probability for coherent recombination of the returning electron prominently visible near the cut-off [Fig.~\ref{fig:multiplot_be}(d)]. \\
%
\begin{figure}
	\centering
	\includegraphics[width=1\linewidth]{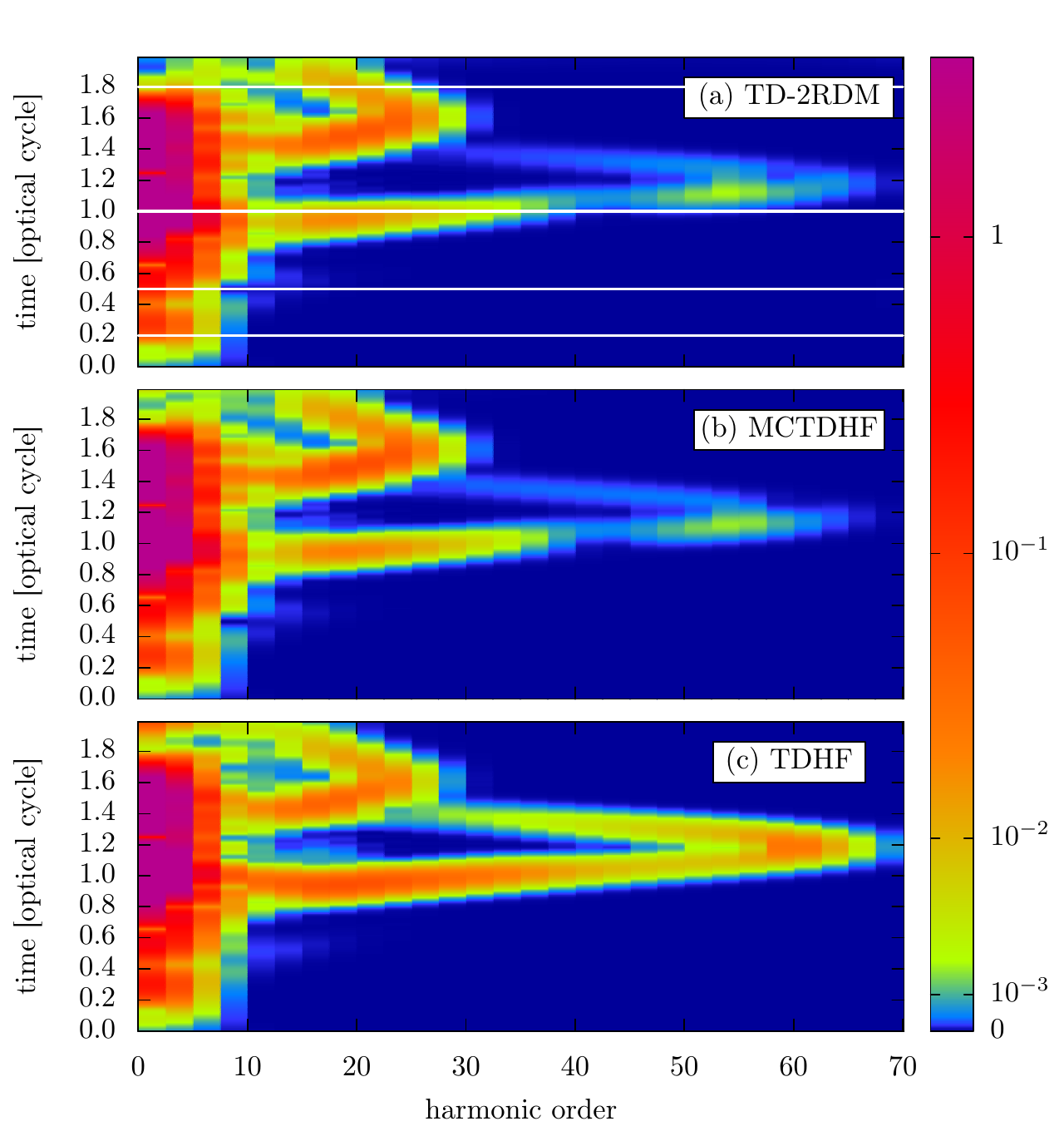}
	\caption{(Color online) Time-frequency spectrum of the emitted high-harmonic radiation from a Be atom subject to a two-cycle laser pulse with $I=4.0\times 10^{14}\rm{W/cm^2}$ and $\lambda = 800\rm nm$ calculated within (a) the TD-2RDM method, (b) MCTDHF and (c) TDHF. The time-frequency spectrum has been obtained by performing a short-time Fourier transformation using a Blackman window function. The intensity of the high-harmonic radiation is shown on a logarithmic scale to highlight the structure of the two recollision events. The four white lines mark times $\tau = 0.2$, $\tau = 0.5$, $\tau = 1.0$ and $\tau = 1.8$ at which snapshots of the particle-hole distribution are taken in Fig.~\ref{fig:ph_correlation}.}
	\label{fig:STFT_be}
\end{figure}\noindent
To trace back the overestimate of the HHG yield predicted by TDHF to the underestimated ionization probability we approximate the intensity of the high-harmonic radiation following the three-step model as 
\begin{align}\label{eq:TSM_intensity}
	I^{\rm HHG}(t_{\rm rec})\propto -\dot N_{\rm ion}(t_{\rm ion})P_{\rm rec}(t_{\rm rec}),
\end{align}
where $\dot N_{\rm ion}(t_{\rm ion})$ is the number of electrons ionized per time interval at ionization time $t_{\rm ion}$ and $P_{\rm rec}(t_{\rm rec})$ is the recombination probability at the recombination time $t_{\rm rec}$. In this classical model the electron is ionized from the atom at time $t_{\rm ion}$ then accelerates in the external laser field before it recombines with the parent ion at time $t_{\rm rec}$. The recombination probability $P_{\rm rec}$ can be approximated by the survival probability, i.e., by the probability of the electron not being ionized. Using both quantities $\dot N^{\rm ion}(t_i)$ and $P^{\rm rec}(t_r)$ from either a MCTDHF or TDHF calculation shows that, indeed, the overestimate of HHG radiation near the cut-off can be reproduced (Fig.~\ref{fig:TSM_be}). \\
The MCTDHF and TD-2RDM method allow for variable occupation numbers of the orbitals permitting the decoupling of single and double ionization. In fact, for the process depicted in Fig.~\ref{fig:multiplot_be}(c,d) the occupation numbers of the natural orbitals change significantly. The accurate reconstruction in combination with the stabilization by purification results in the very good agreement between the TD-2RDM method and MCTDHF for the dipole oscillation as well as the high-harmonic spectrum~[Fig.~\ref{fig:multiplot_be}(c,d)]. \\
To further investigate the structure of the high-harmonic radiation we perform a time-frequency analysis \cite{YakScr03} by applying a short-time Fourier transformation with a Blackman window function \cite{Bla58} using the scaling parameter $\alpha = 0.16$ which features a continuous first derivative allowing a high temporal and spectral resolution. The time-frequency analysis gives us insight at which time high-harmonic radiation of a given frequency is created (Fig.~\ref{fig:STFT_be}). In agreement with the results for the high-harmonic spectrum we find that the TD-2RDM method is capable of almost perfectly reproducing the MCTDHF time-frequency distribution whereas TDHF predicts a significantly stronger intensity at the high-energy cutoff. \\
The structure of the time-frequency behavior of the HHG radiation shows clear signatures of the three-step model~\cite{KraSchKul92,Cor93}. Plotting the recombination energy $E_{\rm rec}=I_p+E_{\rm kin}$ given by ionization plus kinetic energy as a function of time shows two recollision events (Fig.~\ref{fig:TSM_be}). The first event is generated by recombining electrons that have been accelerated in the strong center peak of the two-cycle laser pulse [see inset of Fig.~\ref{fig:multiplot_be}(c)] and the second event by recombining electrons accelerated in the subsequent smaller peak. Obviously, high-harmonic radiation with energy near the cutoff is generated during the first collision. Comparison with the quantum mechanical calculation (Figs.~\ref{fig:STFT_be},\ref{fig:TSM_be}) reveals a small shift to slightly higher cutoff energies in the quantum calculation. This shift is accounted for by the modified cut-off formula predicted by the Lewenstein model \cite{LewBalIva94}. \\
%
\begin{figure}
	\centering
	\includegraphics[width=1\linewidth]{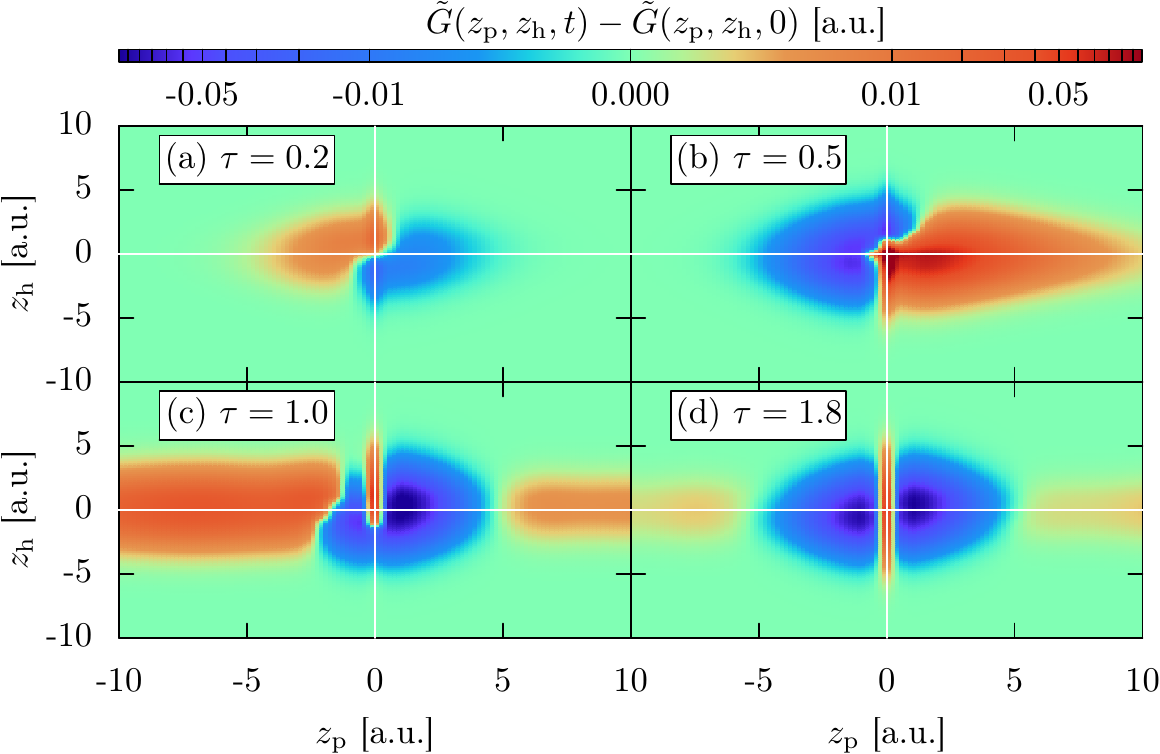}
	\caption{(Color online) Laser field induced deviation from the equilibrium particle-hole distribution function $\tilde{G}^{\rm ph}(z_{\rm p},z_{\rm h},t)~-~\tilde{G}^{\rm ph}(z_{\rm p},z_{\rm h},0)$ [Eq.~(\ref{eq:ph_correlation})] of beryllium for the same laser parameters as in Fig.~\ref{fig:multiplot_be}(c). The times of the snapshots shown are marked in Fig.~\ref{fig:STFT_be} and in the inset of Fig.~\ref{fig:multiplot_be}(c) (note the logarithmic color scale).}
	\label{fig:ph_correlation}
\end{figure}\noindent
%
%
\begin{figure*}[t]
	\centering
	\includegraphics[width=1\linewidth]{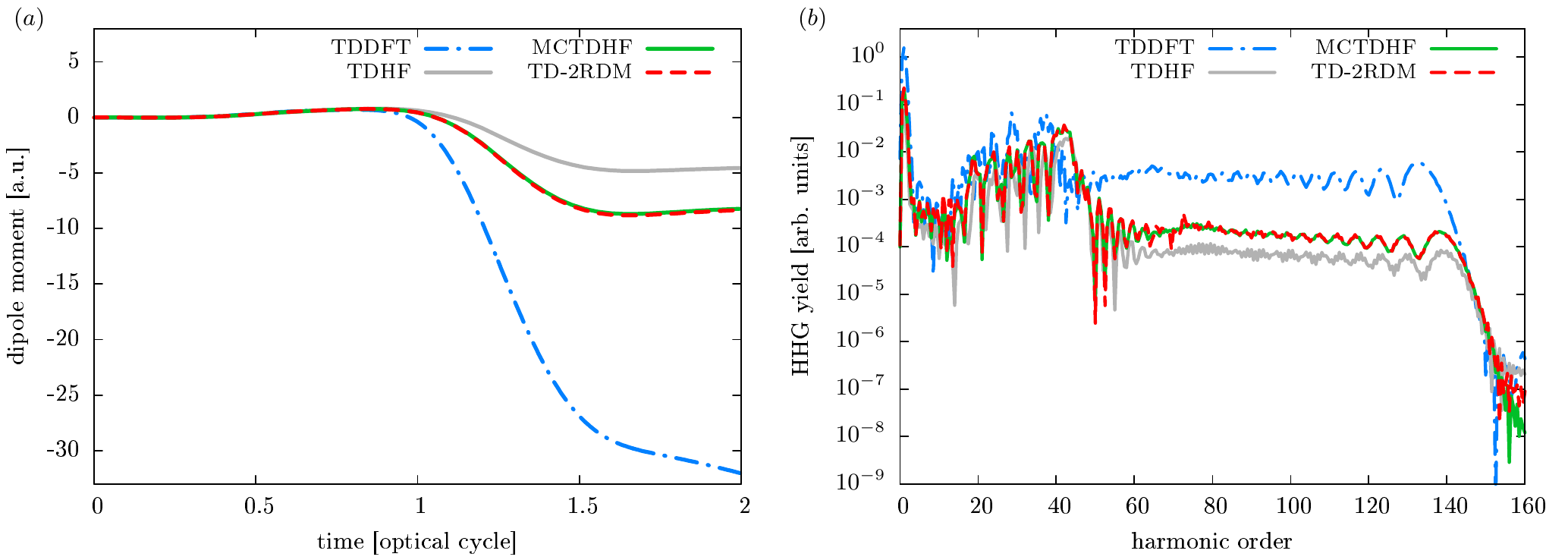}
	\caption{(Color online) (a) Time evolution of the dipole moment and (b) high-harmonic spectrum of Ne subject to a two-cycle laser pulse with $I=10^{15}\rm{W/cm^2}$ and $\lambda = 800\rm nm$. The form of the laser pulse is depicted in the inset of Fig.~\ref{fig:multiplot_be}(c). The TD-2RDM method employing the contraction consistent Nakatsuji-Yasuda reconstruction is compared to MCTDHF, TDHF, and to TDDFT using the local density approximation.}
	\label{fig:multiplot_ne}
\end{figure*}\noindent
%
The present TD-2RDM method allows to explore subtle many-body effects in the HHG process not visible in single-particle or mean-field descriptions. To this end, we analyse the time evolution of the joint particle-hole distribution function 
\begin{align}\label{eq:ph_correlation}
\tilde{G}(z_{\rm p},z_{\rm h})=\langle z_{\rm p}\!\uparrow, z_{\rm h}\!\downarrow \vert \mathcal{K}_2 G_{12} \mathcal{K}_2 \vert z_{\rm p}\!\uparrow, z_{\rm h}\!\downarrow \rangle,
\end{align}
where 
\begin{align}
G_{12}=D_1\delta_2 - D_{12}
\end{align}
is the particle-hole reduced density matrix \cite{Maz07} and 
\begin{align}
\mathcal{K}=\sum_i \vert \phi^g_i \rangle \langle \phi^g_i \vert
\end{align}
is the projection operator onto the initial ground state orbitals $\phi^g_i$. The distribution function $\tilde{G}(z_{\rm p},z_{\rm h})$ measures the probability to find a particle at $z$-coordinate $z_{\rm p}$ while leaving a hole of opposite spin in the ground state with $z$-coordinate $z_{h}$. We show the laser induced particle-hole dynamics by snapshots of the deviation from the equilibrium particle-hole distribution (Fig.~\ref{fig:ph_correlation}) at times marked in Fig.~\ref{fig:STFT_be}(a) and in the inset of Fig.~\ref{fig:multiplot_be}(c). The snapshot at $\tau=0.2$ shows the polarization of the atom under the influence of the first weak peak of the two-cycle laser pulse [see inset of Fig.~\ref{fig:multiplot_be}(c)]. Since the negatively charged particles and the positive holes move under the influence of the electric field in opposite directions, the polarized joint particle-hole distribution is displaced to negative (positive) $z_{\rm p}$ and positive (negative) $z_{\rm h}$. The subsequent peak of the laser pulse at $\tau=0.5$ coincides with the onset of ionization eventually leading to the creation of high-harmonic radiation upon recollision with the parent ion around the time $\tau=1.0$. While the dependence of the joint particle-hole distribution on the particle coordinate closely mirrors the excursion from and return to the core as predicted by the (one-particle) three-step model, the residual hole does not remain frozen but dynamically responds to the motion of the ionized particle and the external field. The hole coordinate performs oscillations phase-shifted by $\pi$ respective to the particle coordinate thereby enhancing the non-linear dipole response of the atom [Fig.~\ref{fig:ph_correlation}(b),(c)]. Near the end of the laser pulse at $\tau=1.8$ the particle-hole distribution deviates from its initial state predominantly due to ionized electrons missing in the the outer shell with $\vert z_{\rm p} \vert>1$. Conversely these ionized electrons have created holes which enhance the particle-hole distribution near the core ($\vert z_{\rm p} \vert<1$).

\subsection{Neon} \label{sec:results/neon}

\begin{figure}
	\centering
	\includegraphics[width=1\linewidth]{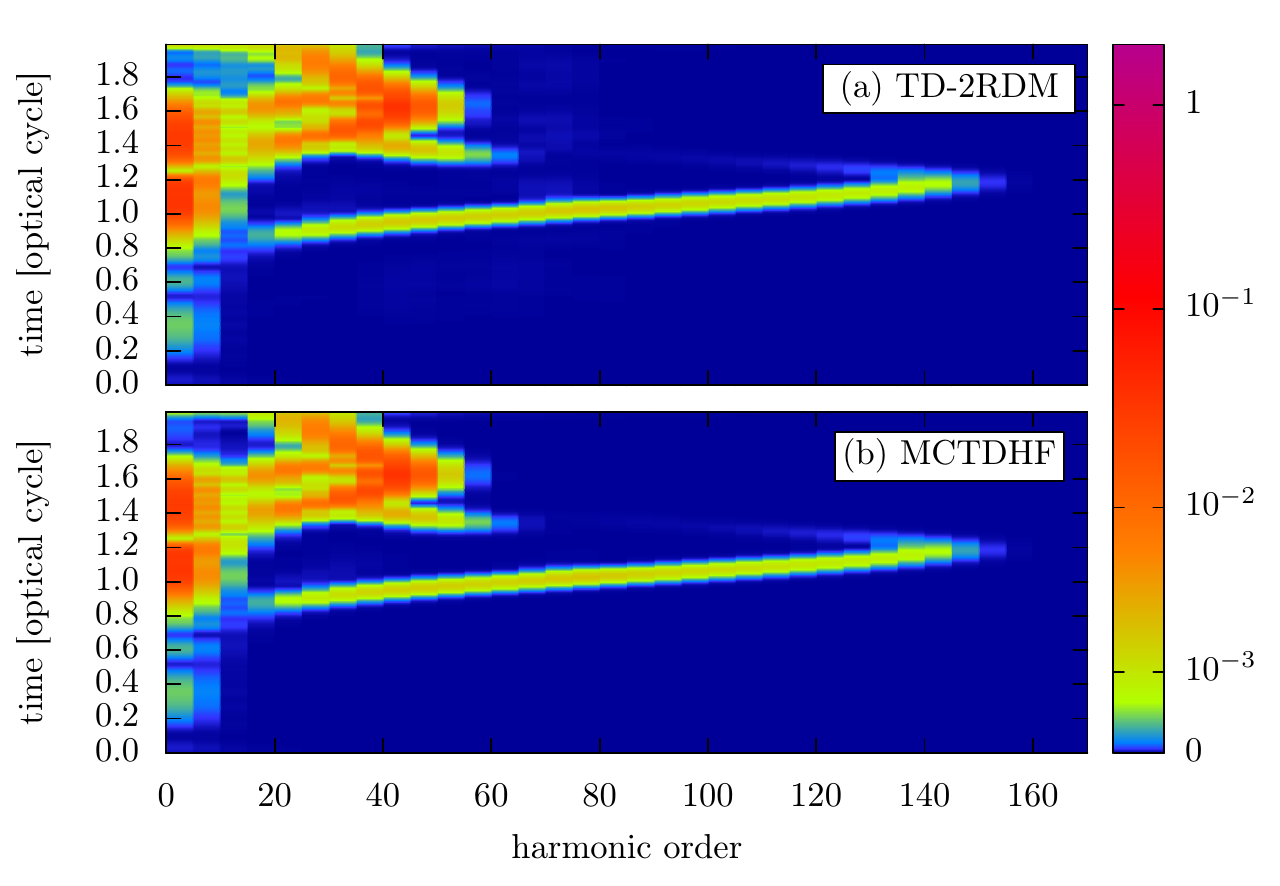}
	\caption{(Color online) Time-frequency spectrum of the emitted high-harmonic radiation of Ne subject to a two-cycle laser pulse with $I=10^{15}\rm{W/cm^2}$ and $\lambda = 800\rm nm$ calculated within (a) the TD-2RDM method and (b) MCTDHF.  The spectrogram has been calculated employing the same method as in Fig.~\ref{fig:STFT_be}.}
	\label{fig:STFT_ne}
\end{figure}

HHG in neon is of great importance as it is used as the workhorse for the generation of attosecond pulses \cite{KraIva09}. Compared to Be the Ne atom has a much larger ionization energy requiring larger laser intensities to obtain a similar amount of ionization and high-harmonic radiation. At the same time, the harmonic cut-off is shifted to much higher frequencies allowing for generation of XUV pulses with energies of up to 100 eV \cite{KraIva09}. Ab-initio descriptions of this process is a considerably more challenging as up to 10 active electrons need to be treated. To simulate HHG in Ne we use a two-cycle laser pulse with $I=10^{15}\rm{W/cm^2}$ leading to a survival probability of about $90\%$ which is significantly larger than $1\%$ for beryllium with $I=4\times 10^{14}\rm{W/cm^2}$ [Fig.~\ref{fig:multiplot_be}(c)]. The small ionization yield for Ne is also reflected in the occupation numbers of the natural orbitals which do not strongly change over time. This allows the present TD-2RDM simulation to be performed even without purification. The shape of the laser pulse is the same as depicted in the inset of Fig.~\ref{fig:multiplot_be}(c).\\
The dipole moment we obtain within the TD-2RDM method is in excellent agreement with the MCTDHF calculation [Fig.~\ref{fig:multiplot_ne}(a)]. Qualitatively, the response of the dipole moment of Ne shows the same slope starting at $\tau \approx 1$ as for the Be atom [Fig.~\ref{fig:multiplot_be}(c)]. This slope signifies the large excursion of the ionized electron driven by the peak electric field near $\tau \approx 1$. The slope in the dipole response and, therefore, ionization is larger for TDDFT and smaller for TDHF as compared to the MCTDHF result [Fig.~\ref{fig:multiplot_ne}(a)]. This is reflected also in the high-harmonic spectrum [Fig.~\ref{fig:multiplot_ne}(b)]. The higher ionization rate of TDDFT leads to a stronger HHG signal compared to MCTDHF and TDHF. In principle, the high-harmonic yield [Eq.~(\ref{eq:TSM_intensity})] also depends on the recombination probability. However, since in the present case ionization is small the dependence on the recombination probability is less important.\\
We find almost perfect agreement between the TD-2RDM method and MCTDHF for the total high-harmonic spectrum  [Fig.~\ref{fig:multiplot_ne}(b)] as well as the time-frequency spectrum (Fig.~\ref{fig:STFT_ne}). Again, the peak structure of the laser pulse is reflected in the two recollision events clearly visible in the time-frequency spectrum. Whereas the first recollision is high in energy its intensity is modest due to the small amount of ionization taking place during the first oscillation of the laser pulse. The second recollision, lower in energy, features a higher intensity due to the strong ionization occurring in the presence of the strong central peak of the laser pulse. 
\section{Conclusions and outlook} {\label{sec:summary}
In this work we have presented the first full 3D simulation of the electronic response of Be and Ne atoms in strong laser pulses using the TD-2RDM method. In particular we have focused on the high-harmonic spectrum obtained from state of the art MCTDHF calculations as an observable to stringently benchmark the accuracy of the TD-2RDM method. The successful application of the TD-2RDM method relies strongly on the reconstruction of the 3RDM entering the closure of the equations of motion for the 2RDM. We compared several reconstructions including second-order reconstructions such as the Nakatsuji-Yasuda \cite{YasNak97} and the Mazziotti reconstruction \cite{Maz00_complete} that are used for the first time within the context of the TD-2RDM method. We found that the Nakatsuji-Yasuda reconstruction in combination with the enforcement of contraction consistency gives the most accurate results for the systems under investigation. In general we found that contraction consistency of the reconstructed 3RDM is not only necessary for the conservation of symmetries in the propagation of the 2RDM but also improves the accuracy of the reconstruction significantly. The results obtained within the TD-2RDM method for the time-dependent atomic polarization as well as for the high-harmonic spectra of Be and Ne are in excellent agreement with MCTDHF reference calculations for a variety of different laser durations and intensities. We have used the present TD-2RDM description to identify the influence of electronic correlations on the HHG process. We investigated the two-particle correlation functions describing the particle-hole dynamics controlling the harmonic emission. These two-point observables are, unlike in mean-field descriptions such as TDHF and TDDFT, directly accessible. The influence of correlation also manifests itself by the significant differences in the HHG spectra calculated by TDHF and TDDFT. In particular, the HHG spectrum of Be in a strongly ionizing laser pulse shows a pronounced overestimation of the HHG yield in the cutoff region if correlation is neglected. This effect can be accurately accounted for within the TD-2RDM method. Furthermore we have calculated the time-frequency spectrum of the emitted high-harmonic radiation. By comparing it to predictions of the three-step model the origin of the correlation induced suppression of the high-harmonic intensity can be delineated. Extensions to larger systems as well as further improvement of the 3RDM reconstruction by including second-order corrections in all elements are planned. 

\section*{Acknowledgments}
We thank Stefan Donsa for helpful discussions on the TD-2RDM method, and Winfried Auzinger, Othmar Koch, and Harald Hofst\"atter for helpful discussions on numerical time propagation..
This research is also supported in part by Grant-in-Aid for Scientific Research (No. 25286064, 26390076, 26600111, and 16H03881) from the Ministry of Education, Culture, Sports, Science and Technology (MEXT) of Japan, and also by the Photon Frontier Net- work Program of MEXT. This research is also partially supported by the Center of Innovation Program from the Japan Science and Technology Agency, JST, and by CREST, JST. The computational results have been achieved using and the Vienna Scientific Cluster (VSC).

\appendix
\section{Spin-block contractions of the 3RDM} \label{sec:appendix/CC}
The formulation of contraction consistency introduced in \ref{sec:methods/reconstruction} [Eq.~(\ref{eq:CC})] is sufficient to ensure energy conservation. For the proper conservation of spin symmetries ($S^2\vert \Psi \rangle=S_z\vert \Psi \rangle=0$) contraction consistency must be extended to relations for the individual 3RDM spin blocks.  In detail the necessary relations are given by
\begin{align} \label{eq:CC_spin}
 	&\sum_m D^{i_1 \uparrow m \uparrow i_2 \downarrow}_{j_1 \uparrow j_2 \uparrow m \downarrow}=D^{i_1 \uparrow i_2 \uparrow}_{j_1 \uparrow j_2 \uparrow} \\
 	&\sum_m D^{i_1 \uparrow i_2 \uparrow m \downarrow}_{j_1 \uparrow m \uparrow j_2 \downarrow}=D^{i_1 \uparrow i_2 \uparrow}_{j_1 \uparrow j_2 \uparrow} \\
 	&\sum_{m}D^{i_1 \uparrow m \uparrow i_2 \downarrow}_{j_1 \uparrow m \uparrow j_2 \downarrow}=\left(\frac{N}{2}-1\right)D^{i_1 \uparrow i_2 \downarrow}_{j_1 \uparrow j_2 \downarrow} \\
 	&\sum_{m}D^{i_1 \uparrow i_2 \uparrow m \downarrow}_{j_1 \uparrow j_2 \uparrow m \downarrow}=\frac{N}{2} D^{i_1 \uparrow i_2 \uparrow}_{j_1 \uparrow j_2 \uparrow}. 
\end{align}
These relations can be derived by using $S^2\vert \Psi \rangle=S_z\vert \Psi \rangle=0$ (see \cite{LacBreSat15} for further details).

\section{Orbital expansion} \label{sec:appendix/orbital_exp}
The numerical implementation of the TD-2RDM method for strong field processes requires the expansion of the 2RDM within a suitable set of $2r$ time-dependent spin orbitals $\phi_{i\sigma}(\mathbf{x},t)=\phi_i(\mathbf{r},t) \otimes \vert \sigma \rangle$,
\begin{align}\label{eq:exp_D_2}
	&D(\mathbf{x_1} \mathbf{x_2};\mathbf{x'_1} \mathbf{x'_2};t)=\nonumber\\
	&\sum_{i_1,i_2,j_1,j_2}D_{j_1 j_2}^{i_1 i_2}(t)\phi_{i_1}(\mathbf{x_1},t) \phi_{i_2}(\mathbf{x_2},t)\phi^*_{j_1}(\mathbf{x'_1},t)\phi^*_{j_2}(\mathbf{x'_2},t),
\end{align}
where we merge the spin $\sigma \in \{\uparrow, \downarrow\}$ and orbital indices $i \in \{1 \dots r\}$. Such an expansion is essential to cover the large simulation box necessary to describe electrons with large excursion radii. For simplicity, we drop here and in the following the labels for the $p$RDM when using the spin-orbital representation, i.e.~$D^{i_1i_2}_{j_1j_2}=[D_{12}]^{i_1i_2}_{j_1j_2}$, $D^{i_1i_2i_3}_{j_1j_2j_3}=[D_{123}]^{i_1i_2i_3}_{j_1j_2j_3}$, since the order $p$ is already uniquely characterized by the orbital-index set.
A convenient choice for the description of the orbital dynamics are the orbital equations of motion of the MCTDHF method
\begin{align} \label{eq:dphi_MCTDH}
	i \partial_t \phi_i(\mathbf{r},t) &=h(\mathbf{r})\phi_i(\mathbf{r},t)+ \hat{Q} \left({\sum_{u} \hat{\Gamma}_u(\mathbf{r},t)[D^{-1}]^u_i} \right),
\end{align}
where 
\begin{align}
	\hat{Q}=1-\sum_{i=1}^{2r} \vert \phi_i \rangle \langle \phi_i \vert
\end{align}
is the orbital projection operator assuring unitary time evolution of the basis orbitals, $[D^{-1}]^u_i$ is the inverse of the 1RDM in the orbital representation, and 
\begin{align}
	\hat{\Gamma}_u(\mathbf{r},t)=\sum_{vwt} D_{u\,t}^{v\,w}\phi_v(\mathbf{r},t)\int \frac{\phi_w(\mathbf{r'},t)\phi^*_t(\mathbf{r'},t)}{|\mathbf{r}-\mathbf{r}'|}\text{d}\mathbf{r'}
\end{align}
represents electron-electron interactions. This last term couples the time evolution of the orbitals to the time evolution of the 2RDM. 
In this work we will focus on closed shell systems (Be and Ne atoms). For these systems the ground state wavefunction is a singlet state and remains, in the absence of spin-orbit interactions, a spin singlet state during propagation. In this case the $(\uparrow \downarrow)$\!\!~-~\!\!block of the 2RDM contains all information on the full 2RDM \cite{LacBreSat15} and the corresponding equation of motion reduces to
\begin{align}\label{eq:eom_approxspin}
	i \partial_t D^{i_1 \uparrow i_2 \downarrow}_{j_1 \uparrow j_2 \downarrow}=&\sum_{k_1,k_2}\big( H^{k_1  k_2 }_{j_1 j_2 }D^{i_1 \uparrow i_2 \downarrow}_{k_1 \uparrow k_2 \downarrow}-D^{k_1 \uparrow k_2 \downarrow}_{j_1 \uparrow j_2\downarrow}H^{i_1 i_2}_{k_1 k_2}\big)\nonumber\\
	+&C^{i_1 \uparrow i_2 \downarrow}_{j_1 \uparrow j_2 \downarrow},
\end{align}
where $H^{k_1 k_2 }_{j_1 j_2 }$ are the matrix elements of the Hamiltonian [Eq.~(\ref{eq:2p_ham})] in the basis of spatial orbitals.
Having to propagate only the $(\uparrow \downarrow)$\!\!~-~\!\!block of the 2RDM significantly reduces the numerical effort because the $(\uparrow \downarrow)$\!\!~-~\!\!block of the collision operator can be written solely in terms of the $(\uparrow \uparrow \downarrow)$\!\!~-~\!\!block of the 3RDM
\begin{align}
	&C^{i_1 \uparrow i_2 \downarrow}_{j_1 \uparrow j_2 \downarrow}=
	I^{i_1 \uparrow i_2 \downarrow}_{j_1 \uparrow j_2 \downarrow}
	+I^{i_2 \uparrow i_1 \downarrow}_{j_2 \uparrow j_1 \downarrow}
	-(I^{j_1 \uparrow j_2 \downarrow}_{i_1 \uparrow i_2 \downarrow}
	+I^{j_2 \uparrow j_1 \downarrow}_{i_2 \uparrow i_1 \downarrow})^*,
	\label{eq:def_Cop_spin}
\end{align}
with
\begin{align}
	I^{i_1 \uparrow i_2 \downarrow}_{j_1 \uparrow j_2 \downarrow}=\sum_{k_1,k_2,k_3}W_{j_1 k_1}^{k_2 k_3}\big( D_{k_2\uparrow k_3\uparrow j_2\downarrow}^{i_1\uparrow k_1\uparrow i_2\downarrow} +D_{j_2\uparrow k_3\uparrow k_2\downarrow}^{i_2\uparrow k_1\uparrow i_1\downarrow}   \big).
	\label{eq:eval_I_spin}
\end{align}

\section{Approximating the TD-2RDM ground state} \label{sec:appendix/init}
The field-free ground states of the MCTDHF and of the TD-2RDM method which serve as initial states of the propagation are idenitical only if the exact collision kernel $C_{12}\{D_{123}\}$ is used in the equation of motion for the 2RDM [Eq.~(\ref{eq:eom_full_d12})]. When the equation of motion is closed by employing an approximate collision kernel $C_{12}\{D^{\rm R}_{123}\}$ based on the reconstruction $D^{\rm R}_{123}$, [Eq.~(\ref{eq:eom_d12})], the exact initial state $D_{12}(t=0)$ is, in general, not a stationary solution of Eq.~(\ref{eq:eom_d12}). Consequently, the admixture of excitations relative to the ground state of Eq.~(\ref{eq:eom_d12}) leads to oscillations and additional artificial "harmonic components" (Fig.~\ref{fig:STFT_ne_not_averaged}). For accurate reconstruction functionals as used here, the deviations from the true ground state and thus the admixture of excitations are small (note the logarithmic scale). The approximation can be further improved upon by projecting out the small admixtures of excitations to the ground state prior to the propagation in the laser field. This is most easily accomplished by field-free propagation [Eq.~(\ref{eq:eom_d12})] of the MCTDHF ground state and performing the time average
\begin{align} \label{eq:average}
 	\bar{D}_{12} = \frac{1}{T} \int_0^T D_{12}(t) \rm{d}t
\end{align}
over a period $T$ sufficiently long compared to the dominant inverse excitation frequencies $T\gg 2\pi / \omega_{\rm excitation}$. This leads to an improved approximation to the proper ground state of the TD-2RDM method and to removal of the residual unphysical oscillations [compare Fig.~\ref{fig:STFT_ne}(a) with Fig.~\ref{fig:STFT_ne_not_averaged}(a)]. In the present case we perform the time average over a time interval of $T=40$ atomic units. Alternatively one could employ well-established algorithms to directly solve the anti-hermitian contracted Schr\"odinger equation \cite{Maz06_antiher}.  
\begin{figure}
	\centering
	\includegraphics[width=1\linewidth]{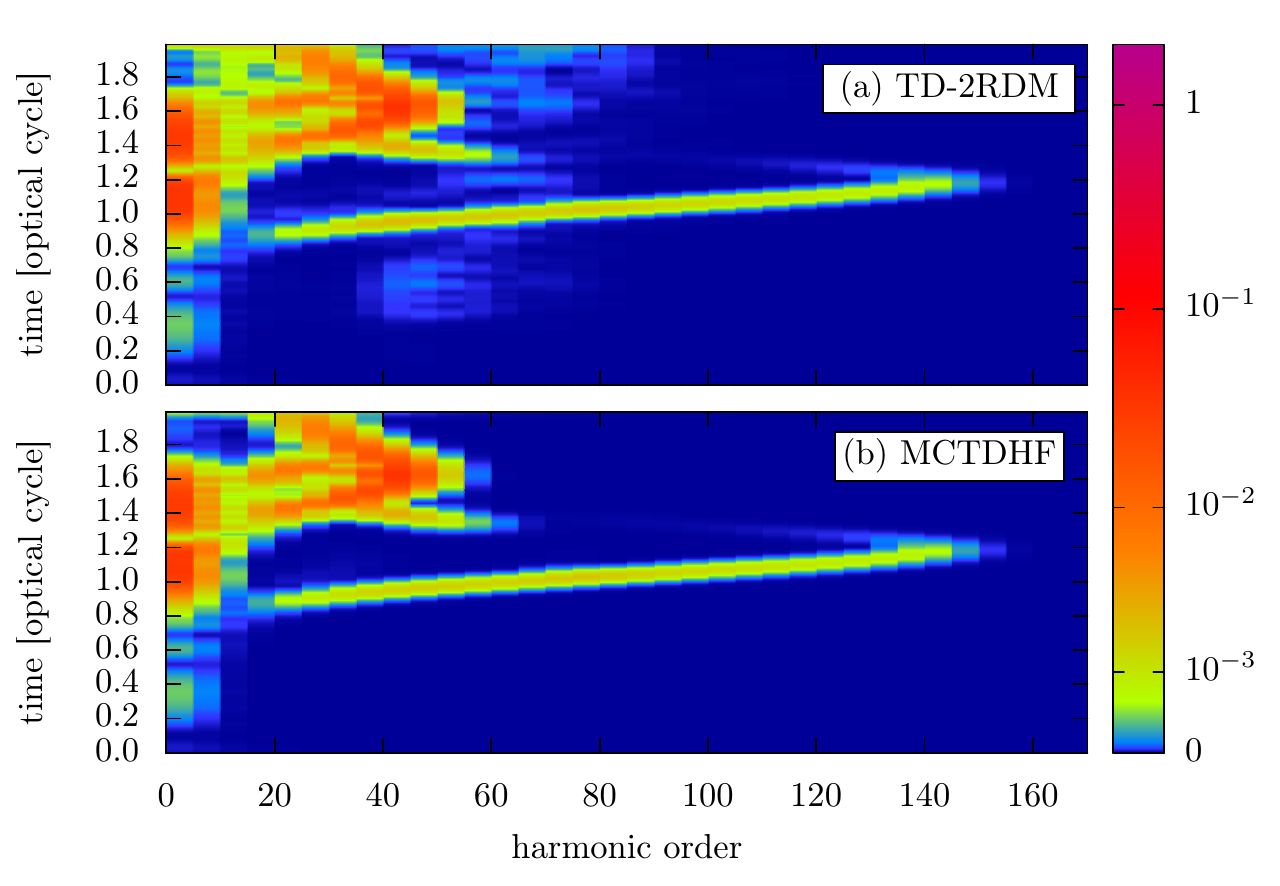}
	\caption{(Color online) Time-frequency spectrum of the emitted high-harmonic radiation of Ne subject to a two-cycle laser pulse with $I=10^{15}\rm{W/cm^2}$ and $\lambda = 800\rm nm$ calculated within (a) the TD-2RDM using the MCTDHF ground state as the initial state and (b) the MCTDHF method. The noise contribution present near the 50$^{\rm th}$ harmonic arise from unphysical excitations present in the "wrong" ground state. The spectrogram has been calculated employing the same method as in Fig.~\ref{fig:STFT_ne}.}
	\label{fig:STFT_ne_not_averaged}
\end{figure}

%
\end{document}